\documentclass{ws-ijmpa}
\usepackage{amsmath}
\def\as{\alpha_s}

\begin{document}

\markboth{Banfi, Laenen}
{Joint resummation for heavy quark production}

%
\catchline{}{}{}{}{}
%

\title{JOINT RESUMMATION FOR HEAVY QUARK PRODUCTION}

\author{\footnotesize ANDREA BANFI}

\address{Cavendish Laboratory, University of Cambridge\\
    Madingley Road, CB3 0HE Cambridge, UK}

\author{ERIC LAENEN\footnote{speaker}}

\address{NIKHEF Theory Group\\
    Kruislaan 409, 1098 SJ Amsterdam, The Netherlands}

\maketitle


\begin{abstract}
  We present the joint threshold and recoil resummed transverse momentum
  distributions for heavy quark hadroproduction, at next-to-leading
  logarithmic accuracy. We exhibit their dependence on the production
  channel and the color configurations, and compare these distributions
to eachother and to NLO.

\end{abstract}

\section{Joint threshold and recoil resummation}    

The formalism \cite{Li:1998is,Laenen:2000ij,Sterman:2004yk} of hadronic cross
sections for the joint resummation of distributions singular at
partonic threshold and at zero recoil has so far been applied to $Z/W$
production \cite{Kulesza:2002rh}, Higgs production \cite{Kulesza:2003wn},
and prompt photon hadroproduction~\cite{Laenen:2000de}.
For the latter $2\rightarrow 2$ process,
the formalism implements the notion that, in
the presence of QCD radiation, the actual transverse momentum produced
by the hard collision is not $\vec{p}_T$ but rather $\vec p_T-\vec
Q_T/2$, with $\vec Q_T$ the total transverse momentum of unobserved soft
recoiling partons. The joint-resummed partonic $p_T$ spectrum
has the form of a hard scattering cross section as a function of
$p_T'\equiv|\vec p_T-\vec Q_T/2|$, convoluted with a {\em
  perturbative}, albeit resummed $\vec Q_T$ distribution.  
We have extended~\cite{Banfi:2004xa} the joint resummation formalism 
to the $p_T$ distribution of
heavy quarks produced in hadronic collisions.  Key differences with
the prompt-photon case are, first, the presence of the heavy quark
mass $m$, preventing a singularity in the hard scattering function
when $Q_T = 2 p_T$ and, second, the possibility of multiple
colored states for the produced heavy quark pair.  

\section{Resummed heavy quark transverse momentum spectra}

We consider the inclusive $p_T$ distribution of a heavy quark produced
via the strong interaction in a hadron-hadron collision at center of
mass (cm) energy $\sqrt{S}$.
Exact higher order corrections to the
differential cross sections for these partonic
processes have been computed to NLO
\cite{Nason:1989zy,Beenakker:1989bq,Beenakker:1991ma,Mangano:1992jk}.
Up to corrections ${\cal O}(1/p^2_T)$,
the observable may at any order \cite{Collins:1989gx} be 
written in the following factorized form 
\begin{equation}
  \label{eq:crs-def}
  \frac{d\sigma_{AB\to Q+X}}{d p_T}=
  \sum_{a,b}\int_0^1 d\xi_a d\xi_b \, \phi_{a/A}(\xi_a,\mu)\phi_{b/B}(\xi_b,\mu)
  \frac{d\hat\sigma_{ab\to Q+X}}{dp_T}(\xi_a,\xi_b,\alpha_s(\mu),p_T)\>,
\end{equation}
with $d\hat\sigma_{ab\to Q+X}/dp_T$ the partonic differential
cross-section, $\phi_{a/A}$ and $\phi_{b/B}$ parton densities, and
$\mu$ the factorization/renormalization scale.  

Threshold enhancements essentially
involve the energy of soft gluons.  In the context of the
factorization (\ref{eq:crs-def}) we define hadronic and partonic
threshold by the conditions $S = 4m_T^2$ and $\hat{s} = 4m_T^2$,
respectively, with $m_T$ the transverse mass $\sqrt{m^2+p_T^2}$.  It
is convenient to define the scaling variables
\begin{equation}
  \label{eq:xT}
  x_T^2=\frac{4 m_T^2}{S}\>,\qquad
  \hat x_T^2=\frac{4 m_T^2}{\xi_a \xi_b S}\>\,,
\end{equation}
so that hadronic (partonic) threshold is at $x_T^2 =1$ ($\hat x_T^2 =
1$).  The higher order corrections to the partonic cross section
$d\hat\sigma_{ab}/dp_T$ contain distributions that are singular at
partonic threshold. Threshold resummation organizes such distributions
to all orders.

There are also recoil effects, resulting from radiation of
soft gluons from initial-state partons.  We wish to treat these
effects in the context of joint threshold and recoil resummation.
We identify a hard scattering with reduced cm
energy squared $Q^2$ and at transverse momentum $\vec{Q}_T$ with
respect to the hadronic cm system. This hard scattering produces a
heavy quark with transverse momentum
\begin{equation}
  \label{eq:transverse}
  \vec{p\,}_T'\equiv\vec p_T-\frac{\vec Q_T}{2}\>.
\end{equation}
The kinematically allowed range for the invariant mass $Q$ of the
heavy quark pair in this hard scattering is limited from below by
$2m_T' = 2\sqrt{m^2 + p_T'^2}$ so that threshold in the context of
joint resummation is defined by
\begin{equation}
  \label{eq:4}
    \tilde x_T^2\equiv\frac{4 m_T'^2}{Q^2} = 1\>.
\end{equation}
A refactorization analysis~\cite{Laenen:2000ij} 
leads to the following expression for the
observable in Eq.~(\ref{eq:crs-def})
\begin{equation}
\label{eq:8}
    \frac{d\sigma_{AB\to Q+X}}{dp_T}
 = \int d^2 Q_T \, \theta(\bar{\mu}-|\vec{Q}_T|)
\frac{d\sigma_{AB\to  Q+X}}{dp_T d^2 \vec Q_T}\,,
\end{equation}
where $\bar{\mu}$ is a cut-off and 
\begin{equation}
\label{eq:2}
  \begin{split}
    &\frac{d\sigma_{AB\to Q+X}}{dp_T d^2 \vec Q_T}= \sum_{ab=q\bar{q},gg}
    p_T \int\frac{d^2 b}{(2\pi)^2}e^{i \vec b \cdot  \vec Q_T}
    \int\frac{dN}{2\pi i} \phi_{a/A}(N,\mu)\>\phi_{b/B}(N,\mu)\>
    e^{E_{ab}(N,b)}\\
    &\frac{e^{-2\>C_F \>t(N)\>(\mathrm{Re} L_\beta+1)}}{4\pi S^2}\left(
      \tilde M^2_{\bf 1}(N)+\tilde M^2_{\bf 8}(N)
    e^{C_A \>t(N)\> \left(\ln \frac{m_T^2}{m^2}+L_\beta\right)}
    \right) \\
    & \qquad\qquad
        \times\left(\frac{S}{4(m^2+|\vec p_T-\vec Q_T/2|^2)}\right)^{N+1}
    \!\!\!\!\>.  
  \end{split}
\end{equation}
Notice in particular the last factor, which provides a kinematic link
between recoil and threshold effects.  The exponential functions
$E_{ab}$ \cite{Laenen:2000ij,Laenen:2000de} are to next-to-leading
logarithmic (NLL) accuracy 
\begin{equation}
  \label{eq:Eab}
  E_{ab}(N,b) = \int_{\chi(N,b)}^Q \frac{d \mu'}{\mu'}
  [A_a(\as(\mu'))+A_b(\as(\mu'))] 2\ln\frac{\bar N \mu'}{Q} -g b^2
  \,,\qquad \bar N = N e^{\gamma_E}\,,
\end{equation}
where the coefficients $A_a$ and $A_b$ 
can be found elsewhere~\cite{Laenen:2000ij}, 
and the function $\chi(N,b)$ is chosen to reproduce either NLL 
resummed recoil or threshold distributions in the appropiate limits
\cite{Kulesza:2002rh}.
We also added to the perturbative exponent 
the non-perturbative (NP) Gaussian smearing term $-gb^2$,
in terms of the impact parameter $b$.
We have introduced the variables
 \begin{equation}
   t(N)=\int_Q^{Q/N}\!\frac{d\mu'}{\mu'}\frac{\alpha_s(\mu')}{\pi}\>,
\;\;
  \mathrm{Re} L_\beta = \frac{1+\beta^2}{2\beta}
  \left(\ln\frac{1-\beta}{1+\beta}\right)
  \>,
  \quad \beta=\sqrt{1-m^2/m_T^2} \>.
\end{equation}
The functions $\tilde M^2_{\bf 1}(N), \tilde M^2_{\bf 8}(N)$ are
the Mellin moments of the lowest order heavy quark production matrix
elements for either the $q\bar{q}$ or $gg$ channel, as appropiate, the index
labeling the color-state of the heavy quark pair. Their explicit expressions
can be found elsewhere\cite{Banfi:2004xa}.
The threshold-resummed result can now easily be derived, by
substituting Eq.~(\ref{eq:2}) into (\ref{eq:8}) 
and neglecting $\vec Q_T$ in the last factor in
Eq.~(\ref{eq:2}). Then the $\vec Q_T$ integral sets $\vec b$ to zero
everywhere, yielding the threshold-resummed result.

\begin{figure}
\centerline{\psfig{file=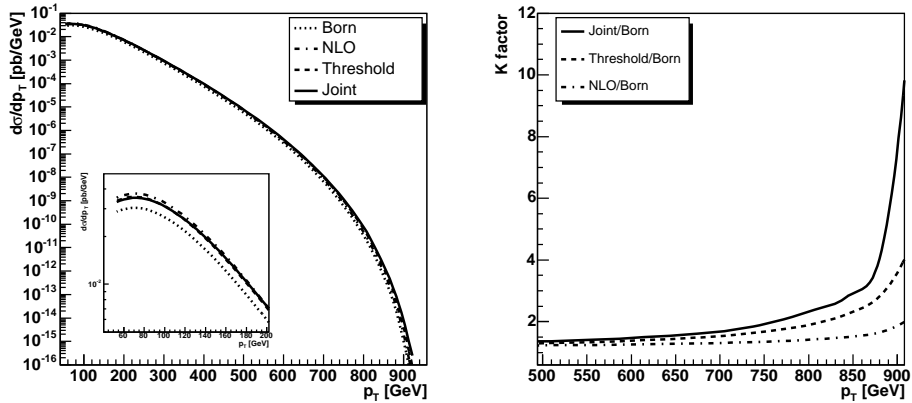,width=\textwidth}}
\vspace*{8pt}
\caption{Top quark $p_T$ spectra and 
    $K$-factors for the $q\bar q$ channel.  \label{fig:ptspectra_qq}}
\end{figure}

To illustrate these analytic results, we show for the 
case of top quark production at the Tevatron
the $p_T$ distribution for the dominant $q\bar q$ channel in 
Fig.~\ref{fig:ptspectra_qq}.
We observe that, while the resummed and NLO curves are close 
for small and moderate
$p_T$ (the inset provides a somewhat better view of the low $p_T$ region), 
for large $p_T$ values the resummed curves depart significantly
from the NLO curve.  Of course, cross sections for top quark
production at such large $p_T$ at the Tevatron are far too small to be
measured, so that our plots at large $p_T$ have only theoretical
interest.  For such large $p_T$ values, the hadronic threshold,
defined in Eq.~\eqref{eq:xT}, approaches the partonic one, where
larger $N$ values dominate, a prerequisite for seeing significant
effects for both resummations.  The enhancements relative to the Born
cross section are shown in the form of a K-factor.
Threshold resummation produces an overall enhancement of the
cross section that increases with increasing $p_T$, yielding e.g. an
enhancement over NLO at $p_T = 800\,\mathrm{GeV}$. Joint resummation almost
doubles that effect: the joint-resummed enhancement at large $p_T$
effectively constitutes a smearing of the threshold-resummed $p_T$ spectrum by
a resummed recoil function.

\section*{Acknowledgments}

This work was supported by the Netherlands
Foundation for Fundamental Research of
Matter (FOM) and the National Organization for Scientific Research
(NWO).


\begin{thebibliography}{0}

\bibitem{Li:1998is}
\protect{H.-n.} Li,
\newblock Phys. Lett. {\bf B454}, 328 (1999), hep-ph/9812363.

\bibitem{Laenen:2000ij}
E.~Laenen, G.~Sterman, and W.~Vogelsang,
\newblock Phys. Rev. {\bf D63}, 114018 (2001), hep-ph/0010080.

\bibitem{Sterman:2004yk}
G.~Sterman and W.~Vogelsang,
\newblock Phys. Rev. {\bf D71}, 114013 (2005), hep-ph/0409234

\bibitem{Kulesza:2002rh}
A.~Kulesza, G.~Sterman, and W.~Vogelsang,
\newblock Phys. Rev. {\bf D66}, 014011 (2002), hep-ph/0202251.

\bibitem{Kulesza:2003wn}
A.~Kulesza, G.~Sterman, and W.~Vogelsang,
\newblock Phys. Rev. {\bf D69}, 014012 (2004), hep-ph/0309264.

\bibitem{Laenen:2000de}
E.~Laenen, G.~Sterman, and W.~Vogelsang,
\newblock Phys. Rev. Lett. {\bf 84}, 4296 (2000), hep-ph/0002078.

\bibitem{Banfi:2004xa}
 A.~Banfi and E.~Laenen,
 Phys.\ Rev.\ D {\bf 71} (2005) 034003,
hep-ph/0411241.

\bibitem{Nason:1989zy}
P.~Nason, S.~Dawson, and R.~K. Ellis,
\newblock Nucl. Phys. {\bf B327}, 49 (1989).

\bibitem{Beenakker:1989bq}
W.~Beenakker, H.~Kuijf, W.~L. van Neerven, and J.~Smith,
\newblock Phys. Rev. {\bf D40}, 54 (1989).

\bibitem{Beenakker:1991ma}
W.~Beenakker, W.~L. van Neerven, R.~Meng, G.~A. Schuler, and J.~Smith,
\newblock Nucl. Phys. {\bf B351}, 507 (1991).

\bibitem{Mangano:1992jk}
M.~L. Mangano, P.~Nason, and G.~Ridolfi,
\newblock Nucl. Phys. {\bf B373}, 295 (1992).

\bibitem{Collins:1989gx}
J.~C. Collins, D.~E. Soper, and G.~Sterman,
\newblock in {\it{Perturbative Quantum Chromodynamics}}, A.H. Mueller ed.,
  World Scientific, Singapore, 1989.



\end{thebibliography}
\end{document}